\documentstyle[12pt]{article}
\input epsf.sty
 1

\catcode`\@=11 
\makeatletter
\def\@seccntformat#1{\csname the#1\endcsname.\hskip 1em}

\makeatother
\begin{document}

\thispagestyle{empty}
\begin{flushright}

{\footnotesize\renewcommand{\baselinestretch}{.75}
           SLAC-PUB-8066\\
           March 1999\\
}
\end{flushright}

\vspace {0.5cm}

\begin{center}
{\large \bf A Preliminary Direct Measurement of the Parity Violating Coupling of 
         the Z$^{0}$ to Strange Quarks, $A_s$$^*$}

\vspace {1.0cm}

 {\bf Hermann Staengle}

\vspace {0.2cm}

Colorado State University, Fort Collins, CO 80523

\vspace {0.6cm}

 {\bf Representing The SLD Collaboration$^{**}$}

\vspace {0.2cm}

Stanford Linear Accelerator Center \\
Stanford University, Stanford, CA~94309 \\

\vspace{1.3cm}
{\bf Abstract}
\end{center}

\renewcommand{\baselinestretch}{1.2}

We present a preliminary direct measurement of the parity violating coupling 
of the $Z^0$ to strange quarks, $A_s$, derived from a sample of approximately
300,000 hadronic decays of Z$^{0}$ bosons produced with a polarized electron 
beam and recorded by the SLD experiment at SLAC between 1993 and 1997. 
$Z^0 \rightarrow s\bar{s}$ events are tagged by the presence in each event 
hemisphere of a high-momentum $K^\pm$, $K_s^0$ or $\Lambda^0$/$\bar\Lambda^0$ 
identified using the Cherenkov Ring Imaging Detector and/or a mass tag.  
The CCD vertex detector is used to suppress the background from heavy flavor
events. The strangeness of the tagged particle is used to sign the event thrust 
axis in the direction of the initial $s$ quark. The coupling $A_s$ is obtained 
directly from a measurement of the left-right-forward-backward production 
asymmetry in polar angle of the tagged $s$ quark.
To reduce the model dependence of the measurement, the background from $u\bar{u}$ 
and $d\bar{d}$ events is measured from the data, as is the analyzing power of 
the method for $s\bar{s}$ events.
We measure:
\begin{eqnarray*}
   A_s = 0.82 \pm 0.10(stat.) \pm 0.08(syst.) (preliminary).
\end{eqnarray*}

\vspace{1.0cm}
\begin{center}
{\it Presented at the American Physical Society (APS) meeting \\
of the Division of Particles and Fields (DPF 99),\\
5--9 January 1999, University of California, Los Angeles.}
\end{center}

\vfil

\noindent
$^*$Work supported in part by Department of Energy contracts 
DE-FG03-93ER40788 and DE-AC03-76SF00515.
\eject

\section{Introduction}           

Measurements of the fermion production asymmetries in the process
$e^+e^- \rightarrow Z^0 \rightarrow f\bar{f}$ provide information on the extent
of parity violation in the coupling of the $Z^0$ boson to fermions of type $f$.
At Born level, the differential production cross section can be expressed in
terms of $x=\cos\theta$, where $\theta$ is the polar angle of the final state
fermion $f$ with respect to the electron beam direction:
\begin{equation}
     \sigma^f(x) =     \frac{d\sigma^f}{dx}
            \propto (1-A_e P_e)(1 + x^2) + 2A_f(A_e-P_e)x ,
\end{equation}
where $P_e$ is the longitudinal polarization of the electron beam, the positron
beam is assumed unpolarized, and the
asymmetry parameters $A_f=2v_f a_f/(v_f^2 + a_f^2)$ are defined in terms of the
vector ($v_f$) and axial-vector ($a_f$) couplings of the $Z^0$ to fermion $f$.
If one measures the polar angle distribution for a given final state $f\bar{f}$, 
one can derive the forward-backward production asymmetry, $A^f_{FB}$, which depends 
on both the initial and final state asymmetry parameters as well as
on the beam polarization:
\begin{equation}
    A^f_{FB}(x) = \frac{\sigma^f(x) - \sigma^f(-x)}{\sigma^f(x) + \sigma^f(-x)} 
                = 2 A_f \frac{A_e - P_e}{1 - A_e P_e} \frac{x}{1+x^2}.
\end{equation}      

For zero polarization, one measures the product of couplings $A_e A_f$.
If one measures the distributions in samples taken with
negative $(L)$ and positive $(R)$ beam polarization of magnitude $P_e$, then one 
can derive the left-right-forward-backward asymmetry, $\tilde{A}^f_{FB}$, which is 
insensitive to the initial state coupling:
\begin{equation} 
\tilde{A}^f_{FB}(x) = \frac{(\sigma^f_L(x) + \sigma^f_R(-x)) - 
                            (\sigma^f_R(x) + \sigma^f_L(-x))}
                           {(\sigma^f_L(x) + \sigma^f_R(-x)) + 
                            (\sigma^f_R(x) + \sigma^f_L(-x))}      
                    = 2 |P_e| A_f \frac{x}{1+x^2}.
\end{equation}    

A number of previous measurements have been made by experiments at LEP and SLC
of $A_e, A_{\mu}, A_{\tau}, A_c$ and $A_b$~\cite{ewlepslc}. 
In contrast, very few measurements exist for the light flavor quarks, due to the
difficulty of tagging specific light flavors.
It has recently been demonstrated experimentally~\cite{lpprl} that light
flavored jets can be tagged by the presence of a high-momentum `leading' identified
particle that has a valence quark of the desired flavor, for example a $K^-$
($K^+$) meson could tag an $s$ ($\bar{s}$) jet.
However the background from other light flavors (a $\bar{u}$ jet can also
produce a leading $K^-$), decays of $B$ and $D$ hadrons, and nonleading kaons
in events of all flavors is large, and neither the signal nor the background
has been well measured experimentally.

The DELPHI collaboration has measured~\cite{delphi}
$A_{FB}^s= 0.131\pm 0.035(stat.)\pm 0.013(syst.)$ and
$A_{FB}^{d,s}=0.112\pm 0.031(stat.)\pm 0.054(syst.)$, respectively. 
However the extraction of the asymmetry parameters from the measured production 
asymmetries is model dependent.
The OPAL collaboration has measured~\cite{opal} 
$A_{FB}^u=0.040\pm 0.067(stat.)\pm 0.028(syst.)$ and
$A_{FB}^{d,s}=0.068\pm 0.035(stat.)\pm 0.011(syst.)$. In their measurement, 
most of the background contributions and analyzing powers are determined 
from double-tagged events in the data.
This eliminates most of the model dependence, but results in limited statistical
precision.

We present a direct measurement of the asymmetry parameter for strange
quarks, $A_s$, using a sample of 300,000 hadronic $Z^0$ decays recorded by the
SLD experiment at the SLAC Linear Collider between 1993 and 1997, with an
average electron beam polarization of 74\%. We use identified strange particles
to tag $s$ and $\bar{s}$ jets. The analyzing power of the tags for true 
$s\bar{s}$ events, as well as the relative contribution of $u \bar u + d \bar d$ 
events were determined from the data.
This procedure removes much of the model dependence, yielding an error that is
statistically dominated.

\section{Hadronic Event and $s\bar{s}$ Event Selection}

A general description of the SLD can be found elsewhere~\cite{sld}.
The trigger and initial selection criteria for hadronic $Z^0$ decays are
described in Ref.~\cite{sldalphas}.
In order to reduce the effects of decays of heavy hadrons, we selected
light flavor events ($u\bar{u}$, $d\bar{d}$ and $s\bar{s}$) by requiring 
each high-quality~\cite{homer} track in the event to have a transverse
impact parameter with respect to the IP of less than three times its estimated
error. Finally, the CRID (Cherenkov Ring Imaging Detector) was required to be 
operational. The selected sample comprised roughly 94,000 events with an estimated
background contribution of 11\% from $c\bar{c}$ events, 2\% from $b\bar{b}$ events, 
and a non-hadronic background contribution of $0.10 \pm 0.05\%$, dominated
by $Z^0 \rightarrow \tau^+\tau^-$ events.

For the purpose of estimating the efficiency and purity of the event
flavor tagging and the particle identification,
we made use of a detailed Monte Carlo (MC) simulation of the detector.
The JETSET 7.4~\cite{jetset} event generator was used, with parameter
values tuned to hadronic $e^+e^-$ annihilation data~\cite{tune},
combined with a simulation of $B$-hadron decays
tuned~\cite{sldsim} to $\Upsilon(4S)$ data and a simulation of the SLD
based on GEANT 3.21~\cite{geant}.

Then high-momentum strange particles are selected.
The CRID allows $K^\pm$ to be separated from $p /\bar p$ and $\pi^\pm$ with 
high purity for tracks with $p > 9$ GeV/c as described in detail in~\cite{bfp}.
A track is tagged as a $K^\pm$ if the log-likelihood for 
this hypothesis exceeds both, the $\pi^\pm$ and the $p/\bar p$
log-likelihoods, by at least 3 units. 
The average purity of the $K^\pm$ sample was estimated using the
simulation to be 89\%. 
$K_s^0$ and $\Lambda^0$/$\bar\Lambda^0$ 
are reconstructed~\cite{bfp} in the modes $K_s^0 \rightarrow \pi^+ \pi^-$ and  
$\Lambda^0 (\bar \Lambda^0) \rightarrow p (\bar p) \pi^\mp$  
and are identified by their long flight distance, reconstructed mass, and accuracy 
of pointing back to the primary interaction point, and are required to have $p > 5$ GeV/c.
Pairs of tracks with invariant mass $m_{\pi\pi}$ within 12 MeV/$c^2$ of 
the nominal $K_s^0$ mass are identified as $K_s^0$.
The simulation predicts that 
the average purity of the $K_s^0$ sample is 91\%.
In the case of the $\Lambda^0$/$\bar\Lambda^0$, information from the 
Cherenkov Ring Imaging Detector is used to identify the $p/ \bar p$ candidate. 
$\Lambda^0$/$\bar\Lambda^0$ are identified by requiring the invariant mass of
pairs of tracks, $m_{p \pi}$,
to be within 5 MeV/$c^2$ of the nominal $\Lambda^0$/$\bar\Lambda^0$ mass and  
a combination of $K_s^0$ rejection and particle identification 
for the high-momentum track.
The simulation predicts that the average purity of the 
$\Lambda^0$/$\bar\Lambda^0$ sample is 84\%.

These strange particles are then used to tag $s$ and $\bar s$ jets as follows.
The event is divided into two hemispheres by a plane perpendicular to the thrust axis.
We require each of the two hemispheres to contain at least one identified 
strange particle ($K^\pm$, $K_s^0$ or $\Lambda^0 / \bar \Lambda^0$); for hemispheres 
with multiple strange particles we only consider the one with the highest momentum.
We require at least one of the two hemispheres to have definite strangeness 
(i.e. to contain a $K^\pm$ or $\Lambda^0 / \bar \Lambda^0$). In events with two 
hemispheres of definite strangeness, the two hemispheres 
are required to have opposite strangeness (e.g. $K^+ K^-$).
This procedure increases the $s\bar{s}$ purity substantially compared with a single 
tag; thus, for these events, the model dependence of the measurement 
is reduced.

Table~\ref{dtags} summarizes the composition of the selected event sample for 
data and simulation for each of the 5 tagging modes used. 
The number of events for each mode shown is in good agreement with the 
MC prediction. The $s \bar s$ purity and $s \bar s$ analyzing power were
estimated from the simulation.

\begin{table}[hbt]
\begin{center}
\caption[Summary of selected event sample]{\label{dtags} 
         Summary of selected event sample for 5 modes in data and simulation.}
\bigskip
\begin{tabular}{|c||c|c|c|c|} \hline 
Mode & Data & MC prediction & $s \bar s$ purity & $s \bar s$ analyzing power \\
\hline \hline

$K^+ K^-$                               & 619  & 620  & 0.76 & 0.94  \\
\hline
$K^+ \Lambda^0, K^- \bar \Lambda^0$     &  86  &  82  & 0.65 & 0.86  \\
\hline
$\Lambda^0 \bar \Lambda^0$              &   1  &   6  & 0.59 &       \\
\hline
$K^\pm K_s^0$                           & 502  & 531  & 0.64 & 0.68  \\
\hline
$\Lambda^0 K_s^0, \bar \Lambda^0 K_s^0$ &  52  &  52  & 0.54 & 0.44  \\

\hline  \hline
Total:                                  & 1260 & 1291 & 0.69 & 0.82 \\

\hline
\end{tabular}
\end{center}
\end{table}

The combined $s\bar{s}$ purity of all modes is 69\%, and the predicted 
background in the selected event sample consists of 10\% $u \bar u$, 9\% $d \bar d$, 
11\% $c \bar c$, and 1\% $b \bar b$ events.
The analyzing power is defined as
$a_s = \frac{N^{right}_s - N^{wrong}_s}{N^{right}_s + N^{wrong}_s}$  
where $N^{right}_s (N^{wrong}_s)$ denotes the number of $s\bar s$ events
in which a particle of negative strangeness is found in the true $s (\bar s)$ hemisphere.
The average analyzing power for all modes is predicted by the simulation to be 0.82. 
The $K^\pm K^\mp$ mode has a substantially higher analyzing power and $s\bar{s}$ purity
than the other modes.
The initial $s$ quark direction is approximated by the thrust axis, $\hat t$ of 
the event, signed to point in the direction of negative strangeness,
$x = cos \theta_s = S  \frac{\vec{p} \cdot \hat{t}}{|\vec{p} \cdot \hat{t}|}  \hat t_z$,
where $S$ and $\vec{p}$ denote the strangeness and the momentum of the tagging particle.

\begin{figure}[htb]
 \centering
 \epsfxsize12cm
 \epsfxsize12cm
 \leavevmode
 \epsffile{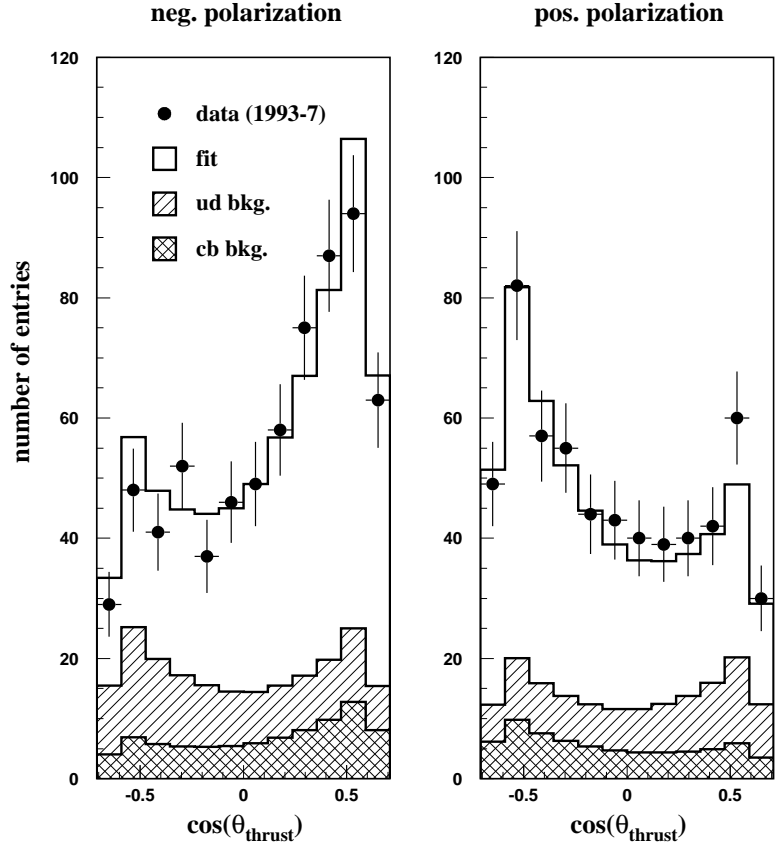}
 \caption{Polar angle distributions of the thrust axis, signed to point in the direction
          of negative strangeness, 
          of the tagged strange particle, for negative (left) and positive (right) beam
          polarization. The dots show data and the histogram shows our fit to the data.
          Our estimates of the non-$s \bar s$ backgrounds are indicated by the hatched
          histograms.}
 \label{asymplot}
 \end{figure}

Figure~\ref{asymplot} shows the polar angle distributions, for all 5 tagging modes 
combined, of the signed thrust axis for left-handed ($P_e < 0$) and right-handed 
($P_e > 0$) electron beams. The expected production asymmetries, of 
opposite sign for the left-handed and the right-handed beams, are clearly visible.

\section{Extraction of $A_s$ and Systematic Uncertainties}

$A_s$ is extracted from these distributions by a binned maximum likelihood fit. 
The fitting function is given by:
\begin{equation}
    P(x) = D(x) \sum_{f}^{} {N_f (1 + x^2 + 2 (1 + \delta) a_f A_f A_Z x}).
\end{equation} 
The function $D(x)$ describes the acceptance and the 
strange particle identification efficiencies. 
$N_f = N_{events} R_f \epsilon_f$ denotes the number of events in the sample of flavor
$f$ ($f = u,d,s,c,b$) in terms of the number of selected hadronic events $N_{events}$, 
$R_f = \Gamma(Z^0 \rightarrow f \bar f)/\Gamma(Z^0 \rightarrow hadrons)$
and the tagging efficiencies $\epsilon_f$; 
$\delta = -0.013$ corrects for the effects of hard gluon radiation~\cite{stavolsen};
$a_f$ denotes the analyzing power for tagging the $f$ rather than the 
$\bar f$ direction; 
$A_f$ is the asymmetry parameter for flavor $f$; 
and $A_Z = (A_e - P_e)/(1 - A_e P_e)$.
The function $D(x)$ was calculated from the simulation and verified 
by comparing data and simulated $x$-distributions of identified $K^\pm$, $K_s^0$,
and $\Lambda^0 /\bar \Lambda^0$. 
The parameters $\epsilon_c$, $\epsilon_b$, and $a_c$, $a_b$ for the heavy flavors are 
taken from the MC simulation~\cite{sldsim}.
The world average experimental measurements of the parameters $A_c$, $A_b$, $R_c$, 
$R_b$~\cite{ewlepslc} were used.
For the light flavors, the relevant parameters in the fitting function are derived 
where possible from the data as described below.
The total number of light flavor events, $N_{uds}$, 
is determined by subtracting the simulated number of heavy flavor events
from the entire event sample. 
The ratio $N_{ud}/N_s$ is $0.27 \pm 0.03$.
The asymmetry parameters $A_u$ and $A_d$ are set to the Standard Model values.
The $s\bar s$ analyzing power, $a_s$, is $0.82 \pm 0.03$ (averaged over all modes).
The combined $(u\bar u + d\bar d)$ analyzing power, $a_{ud}$, is estimated 
to be $-0.41 \pm 0.24$. 
The result of the fit is shown as a histogram in Figure~\ref{asymplot}. The fit quality 
is good with a $\chi^2$ of 12.9 for 24 bins. 
Also included are our estimates of non-$s \bar s$ background. 
The cross-hatched histograms indicate $c\bar c + b \bar b$ backgrounds which are 
seen to show asymmetries of the same sign and similar slope to the total distribution.
The hatched histograms indicate $u\bar u + d \bar d$ backgrounds showing 
asymmetries of the opposite sign and slope to the total distribution. 
The $A_s$ value extracted from the fit is $A_s = 0.82 \pm 0.10(stat.)$.

The understanding of the parameters used as inputs to the fitting function and
of their uncertainties is crucial to this analysis.
The characteristics of heavy flavor events relevant to this analysis have been
measured experimentally, and our simulation~\cite{jetset,tune,geant} has been 
tuned~\cite{sldsim} to reproduce these results.
Standard systematic variations of the heavy flavor simulation
were considered, and the resulting uncertainties are a small 
contribution to the total systematic error.
Other small contributions to the systematic error include those from the 0.6\% 
uncertainty in the correction for the effect of hard gluon radiation, and 
the 0.8\% uncertainty in the beam polarization.
For the light flavors, there are few experimental constraints on the relevant 
input parameters.
Qualitative features such as leading particle production~\cite{lpprl}, short range
rapidity correlations between high-momentum $K K$ and baryon-antibaryon 
pairs~\cite{davecorrel} and long-range correlations between 
several particle species~\cite{davecorrel} have been observed 
experimentally, but these results are not sufficient to quantify the  
analyzing power of the strange-particle tag or the $u \bar u$ and $d \bar d$ background. 
Our MC simulation provides a reasonable description of the above observations
and was used to evaluate the parameters used in the fit, but was
calibrated using data.
For the analyzing power in $s \bar{s}$ events,
we investigated the rate of production of wrong-sign kaons by 
counting events in which we find three identified $K^{\pm}$ and/or $K_s^0$.
Since we found that the MC prediction for the number of 3-kaon jets 
is consistent with the data, we used the simulated 
$a_s = 0.82$ as our central value for the analyzing power in $s\bar{s}$ events
and the statistical error on the data-MC comparison as uncertainty.
We also counted hemispheres containing a $K^+ K^+$ or $K^- K^-$ pair, obtaining 
consistent but less precise constraints.

\begin{table}[hbt]
\begin{center}
\caption[Summary systematic uncertainties]{\label{systerr} Summary of systematic uncertainties.}
\bigskip
\begin{tabular}{|c||c|c|c|} \hline
 Source & Comments & Systematic variation & $\delta A_s / A_s$ \\
\hline \hline
           
heavy flavor modelling   & MC/world averages    & Ref.~\cite{ewlepslc,tune,sldsim} & $0.008$ \\
\hline 
hard gluon radiation     & Stav-Olsen with      & $(1.3 \pm 0.6)\%$  & $0.006$ \\
                         & bias correction      &                    &         \\
\hline 
beam polarization        & data                 & $(73.7\pm 0.8)\%$  & $0.008$ \\
\hline 
$a_s$                    & MC constrained by    & $0.82 \pm 0.03$    & $0.038$ \\
                         & 3-K jets in data     &                    &         \\
\hline
$a_{ud}$                 & $-a_s < a_{ud} < 0$  & $-0.41 \pm 0.24$   & $0.071$ \\
$A_{ud}$                 & Standard Model       &  --                & --      \\
\hline  
$N_{ud} / N_s$           & MC constrained by    & $0.27 \pm 0.03$    & $0.046$ \\
                         & 2-K jets in data     &                    &         \\
\hline 
MC statistics            &                      &                    & $0.021$ \\
\hline \hline
Total:                   &                      &                    & $0.096$ \\

\hline
\end{tabular}
\end{center}
\end{table}

For the calibration of the relative $u \bar u + d \bar d$ background level, $N_{ud}/N_s$,
we counted the number of hemispheres containing an identified $K^+$-$K^-$
pair or an identified $K^\pm$-$K^0$ pair. The MC prediction
is consistent with the data and was used as the central value.
Another, less precise contraint was obtained from events that were tagged by 
kaons of the same sign in both hemispheres.
The above checks are also sensitive to the analyzing power of 
$u\bar u + d\bar d$ events, $a_{ud}$.
However, with the present event statistics we cannot obtain a tight constraint on 
this quantity.  We therefore assume that $a_{ud}$ must be negative, since $u$ and $d$ 
jets must produce a leading $K^+$ rather than $K^-$, and that $|a_{ud}|$ must be less 
than $a_s$, since there is always a companion particle of opposite strangeness in 
a $u$ or $d$ jet that will tend to dilute the analyzing power. We take these as 
hard limits, $-0.82 < a_{ud} < 0$, use the middle of the range for our central value 
and assign an uncertainty equal to the range divided by $\sqrt{12}$. The simulation 
predicts a value of $a_{ud}=-0.38$, consistent with our estimate. 
Table~\ref{systerr} summarizes the systematic uncertainties.
The total systematic uncertainty is $\delta A_s = 0.08$.

\section{Summary and Conclusion}

We have presented a preliminary direct measurement of the parity violating coupling 
of the $Z^0$ to strange quarks, $A_s$, derived from a sample of approximately 300,000 
hadronic decays of Z$^{0}$ bosons produced with a polarized electron beam and 
recorded by the SLD experiment at SLAC between 1993 and 1997.
The coupling $A_s$ is obtained directly from a measurement of the 
left-right-forward-backward production asymmetry in polar angle of the tagged 
$s$ quark.
The background from $u\bar{u}$ and $d\bar{d}$ events is measured from the 
data, as is the analyzing power of the method for $s\bar{s}$ events.
A binned maximum likelihood fit is used to obtain the result:  
\begin{equation}
    A_s  = 0.82 \pm 0.10(stat.) \pm 0.08 (syst.) (preliminary).
\end{equation}    

\noindent
This result is consistent with the Standard Model expectation for $A_s$.
Our measurement can be used to test the universality of the coupling constants
by comparing it with the world average value for $A_b$~\cite{ewlepslc}. 
The two measurements are consistent.

In order to compare with previous measurements of $A_{FB}^{s}$ and $A_{FB}^{d,s}$
(see section 1), we must assume a value of $A_e$. Using $A_e = 0.1499$~\cite{ewlepslc}
and neglecting the small uncertainty on $A_e$, 
the DELPHI measurements translate into
$A_s = 1.165 \pm 0.311(stat.) \pm 0.116(syst.)$
and
$A_{d,s} = 0.996 \pm 0.276(stat.) \pm 0.480(syst.)$. 
Similarly, the OPAL measurement yields
$A_{d,s} = 0.605 \pm 0.311(stat.) \pm 0.098(syst.)$.
Our measurement is consistent with these and represents a substantial
improvement in precision.

%


\section*{$^{**}$List of Authors} 
%
%
%
\begin{center}
\def\iADEL{$^{(1)}$}
\def\iAOMORI{$^{(2)}$}
\def\iBOLO{$^{(3)}$}
\def\iBRUN{$^{(4)}$}
\def\iBU{$^{(5)}$}
\def\iCINC{$^{(6)}$}
\def\iCOLO{$^{(7)}$}
\def\iCOLU{$^{(8)}$}
\def\iCSU{$^{(9)}$}
\def\iFERR{$^{(10)}$}
\def\iFRAS{$^{(11)}$}
\def\iILLI{$^{(12)}$}
\def\iLBL{$^{(13)}$}
\def\iLTU{$^{(14)}$}
\def\iMASS{$^{(15)}$}
\def\iMISSI{$^{(16)}$}
\def\iMIT{$^{(17)}$}
\def\iMOSCOW{$^{(18)}$}
\def\iNAGO{$^{(19)}$}
\def\iOREG{$^{(20)}$}
\def\iOXF{$^{(21)}$}
\def\iPADO{$^{(22)}$}
\def\iPERU{$^{(23)}$}
\def\iPISA{$^{(24)}$}
\def\iRAL{$^{(25)}$}
\def\iRUTG{$^{(26)}$}
\def\iSLAC{$^{(27)}$}
\def\iSOGA{$^{(28)}$}
\def\iSOONG{$^{(29)}$}
\def\iTENN{$^{(30)}$}
\def\iTOHO{$^{(31)}$}
\def\iUCSB{$^{(32)}$}
\def\iUCSC{$^{(33)}$}
\def\iVAND{$^{(34)}$}
\def\iWASH{$^{(35)}$}
\def\iWISC{$^{(36)}$}
\def\iYALE{$^{(37)}$}

  \baselineskip=.75\baselineskip  
\mbox{K. Abe\unskip,\iAOMORI}
\mbox{K.  Abe\unskip,\iNAGO}
\mbox{T. Abe\unskip,\iSLAC}
\mbox{I.Adam\unskip,\iSLAC}
\mbox{T.  Akagi\unskip,\iSLAC}
\mbox{N. J. Allen\unskip,\iBRUN}
\mbox{A. Arodzero\unskip,\iOREG}
\mbox{W.W. Ash\unskip,\iSLAC}
\mbox{D. Aston\unskip,\iSLAC}
\mbox{K.G. Baird\unskip,\iMASS}
\mbox{C. Baltay\unskip,\iYALE}
\mbox{H.R. Band\unskip,\iWISC}
\mbox{M.B. Barakat\unskip,\iLTU}
\mbox{O. Bardon\unskip,\iMIT}
\mbox{T.L. Barklow\unskip,\iSLAC}
\mbox{J.M. Bauer\unskip,\iMISSI}
\mbox{G. Bellodi\unskip,\iOXF}
\mbox{R. Ben-David\unskip,\iYALE}
\mbox{A.C. Benvenuti\unskip,\iBOLO}
\mbox{G.M. Bilei\unskip,\iPERU}
\mbox{D. Bisello\unskip,\iPADO}
\mbox{G. Blaylock\unskip,\iMASS}
\mbox{J.R. Bogart\unskip,\iSLAC}
\mbox{B. Bolen\unskip,\iMISSI}
\mbox{G.R. Bower\unskip,\iSLAC}
\mbox{J. E. Brau\unskip,\iOREG}
\mbox{M. Breidenbach\unskip,\iSLAC}
\mbox{W.M. Bugg\unskip,\iTENN}
\mbox{D. Burke\unskip,\iSLAC}
\mbox{T.H. Burnett\unskip,\iWASH}
\mbox{P.N. Burrows\unskip,\iOXF}
\mbox{A. Calcaterra\unskip,\iFRAS}
\mbox{D.O. Caldwell\unskip,\iUCSB}
\mbox{D. Calloway\unskip,\iSLAC}
\mbox{B. Camanzi\unskip,\iFERR}
\mbox{M. Carpinelli\unskip,\iPISA}
\mbox{R. Cassell\unskip,\iSLAC}
\mbox{R. Castaldi\unskip,\iPISA}
\mbox{A. Castro\unskip,\iPADO}
\mbox{M. Cavalli-Sforza\unskip,\iUCSC}
\mbox{A. Chou\unskip,\iSLAC}
\mbox{E. Church\unskip,\iWASH}
\mbox{H.O. Cohn\unskip,\iTENN}
\mbox{J.A. Coller\unskip,\iBU}
\mbox{M.R. Convery\unskip,\iSLAC}
\mbox{V. Cook\unskip,\iWASH}
\mbox{R. Cotton\unskip,\iBRUN}
\mbox{R.F. Cowan\unskip,\iMIT}
\mbox{D.G. Coyne\unskip,\iUCSC}
\mbox{G. Crawford\unskip,\iSLAC}
\mbox{C.J.S. Damerell\unskip,\iRAL}
\mbox{M. N. Danielson\unskip,\iCOLO}
\mbox{M. Daoudi\unskip,\iSLAC}
\mbox{N. de Groot\unskip,\iSLAC}
\mbox{R. Dell'Orso\unskip,\iPERU}
\mbox{P.J. Dervan\unskip,\iBRUN}
\mbox{R. de Sangro\unskip,\iFRAS}
\mbox{M. Dima\unskip,\iCSU}
\mbox{A. D'Oliveira\unskip,\iCINC}
\mbox{D.N. Dong\unskip,\iMIT}
\mbox{P.Y.C. Du\unskip,\iTENN}
\mbox{R. Dubois\unskip,\iSLAC}
\mbox{B.I. Eisenstein\unskip,\iILLI}
\mbox{V. Eschenburg\unskip,\iMISSI}
\mbox{E. Etzion\unskip,\iWISC}
\mbox{S. Fahey\unskip,\iCOLO}
\mbox{D. Falciai\unskip,\iFRAS}
\mbox{C. Fan\unskip,\iCOLO}
\mbox{J.P. Fernandez\unskip,\iUCSC}
\mbox{M.J. Fero\unskip,\iMIT}
\mbox{K.Flood\unskip,\iMASS}
\mbox{R. Frey\unskip,\iOREG}
\mbox{T. Gillman\unskip,\iRAL}
\mbox{G. Gladding\unskip,\iILLI}
\mbox{S. Gonzalez\unskip,\iMIT}
\mbox{E.L. Hart\unskip,\iTENN}
\mbox{J.L. Harton\unskip,\iCSU}
\mbox{A. Hasan\unskip,\iBRUN}
\mbox{K. Hasuko\unskip,\iTOHO}
\mbox{S. J. Hedges\unskip,\iBU}
\mbox{S.S. Hertzbach\unskip,\iMASS}
\mbox{M.D. Hildreth\unskip,\iSLAC}
\mbox{J. Huber\unskip,\iOREG}
\mbox{M.E. Huffer\unskip,\iSLAC}
\mbox{E.W. Hughes\unskip,\iSLAC}
\mbox{X.Huynh\unskip,\iSLAC}
\mbox{H. Hwang\unskip,\iOREG}
\mbox{M. Iwasaki\unskip,\iOREG}
\mbox{D. J. Jackson\unskip,\iRAL}
\mbox{P. Jacques\unskip,\iRUTG}
\mbox{J.A. Jaros\unskip,\iSLAC}
\mbox{Z.Y. Jiang\unskip,\iSLAC}
\mbox{A.S. Johnson\unskip,\iSLAC}
\mbox{J.R. Johnson\unskip,\iWISC}
\mbox{R.A. Johnson\unskip,\iCINC}
\mbox{T. Junk\unskip,\iSLAC}
\mbox{R. Kajikawa\unskip,\iNAGO}
\mbox{M. Kalelkar\unskip,\iRUTG}
\mbox{Y. Kamyshkov\unskip,\iTENN}
\mbox{H.J. Kang\unskip,\iRUTG}
\mbox{I. Karliner\unskip,\iILLI}
\mbox{H. Kawahara\unskip,\iSLAC}
\mbox{Y. D. Kim\unskip,\iSOGA}
\mbox{R. King\unskip,\iSLAC}
\mbox{M.E. King\unskip,\iSLAC}
\mbox{R.R. Kofler\unskip,\iMASS}
\mbox{N.M. Krishna\unskip,\iCOLO}
\mbox{R.S. Kroeger\unskip,\iMISSI}
\mbox{M. Langston\unskip,\iOREG}
\mbox{A. Lath\unskip,\iMIT}
\mbox{D.W.G. Leith\unskip,\iSLAC}
\mbox{V. Lia\unskip,\iMIT}
\mbox{C.-J. S. Lin\unskip,\iSLAC}
\mbox{X. Liu\unskip,\iUCSC}
\mbox{M.X. Liu\unskip,\iYALE}
\mbox{M. Loreti\unskip,\iPADO}
\mbox{A. Lu\unskip,\iUCSB}
\mbox{H.L. Lynch\unskip,\iSLAC}
\mbox{J. Ma\unskip,\iWASH}
\mbox{G. Mancinelli\unskip,\iRUTG}
\mbox{S. Manly\unskip,\iYALE}
\mbox{G. Mantovani\unskip,\iPERU}
\mbox{T.W. Markiewicz\unskip,\iSLAC}
\mbox{T. Maruyama\unskip,\iSLAC}
\mbox{H. Masuda\unskip,\iSLAC}
\mbox{E. Mazzucato\unskip,\iFERR}
\mbox{A.K. McKemey\unskip,\iBRUN}
\mbox{B.T. Meadows\unskip,\iCINC}
\mbox{G. Menegatti\unskip,\iFERR}
\mbox{R. Messner\unskip,\iSLAC}
\mbox{P.M. Mockett\unskip,\iWASH}
\mbox{K.C. Moffeit\unskip,\iSLAC}
\mbox{T.B. Moore\unskip,\iYALE}
\mbox{M.Morii\unskip,\iSLAC}
\mbox{D. Muller\unskip,\iSLAC}
\mbox{V.Murzin\unskip,\iMOSCOW}
\mbox{T. Nagamine\unskip,\iTOHO}
\mbox{S. Narita\unskip,\iTOHO}
\mbox{U. Nauenberg\unskip,\iCOLO}
\mbox{H. Neal\unskip,\iSLAC}
\mbox{M. Nussbaum\unskip,\iCINC}
\mbox{N.Oishi\unskip,\iNAGO}
\mbox{D. Onoprienko\unskip,\iTENN}
\mbox{L.S. Osborne\unskip,\iMIT}
\mbox{R.S. Panvini\unskip,\iVAND}
\mbox{H. Park\unskip,\iOREG}
\mbox{C. H. Park\unskip,\iSOONG}
\mbox{T.J. Pavel\unskip,\iSLAC}
\mbox{I. Peruzzi\unskip,\iFRAS}
\mbox{M. Piccolo\unskip,\iFRAS}
\mbox{L. Piemontese\unskip,\iFERR}
\mbox{E. Pieroni\unskip,\iPISA}
\mbox{K.T. Pitts\unskip,\iOREG}
\mbox{R.J. Plano\unskip,\iRUTG}
\mbox{R. Prepost\unskip,\iWISC}
\mbox{C.Y. Prescott\unskip,\iSLAC}
\mbox{G.D. Punkar\unskip,\iSLAC}
\mbox{J. Quigley\unskip,\iMIT}
\mbox{B.N. Ratcliff\unskip,\iSLAC}
\mbox{T.W. Reeves\unskip,\iVAND}
\mbox{J. Reidy\unskip,\iMISSI}
\mbox{P.L. Reinertsen\unskip,\iUCSC}
\mbox{P.E. Rensing\unskip,\iSLAC}
\mbox{L.S. Rochester\unskip,\iSLAC}
\mbox{P.C. Rowson\unskip,\iCOLU}
\mbox{J.J. Russell\unskip,\iSLAC}
\mbox{O.H. Saxton\unskip,\iSLAC}
\mbox{T. Schalk\unskip,\iUCSC}
\mbox{R.H. Schindler\unskip,\iSLAC}
\mbox{B.A. Schumm\unskip,\iUCSC}
\mbox{J. Schwiening\unskip,\iSLAC}
\mbox{S. Sen\unskip,\iYALE}
\mbox{V.V. Serbo\unskip,\iWISC}
\mbox{M.H. Shaevitz\unskip,\iCOLU}
\mbox{J.T. Shank\unskip,\iBU}
\mbox{G. Shapiro\unskip,\iLBL}
\mbox{D.J. Sherden\unskip,\iSLAC}
\mbox{K. D. Shmakov\unskip,\iTENN}
\mbox{C. Simopoulos\unskip,\iSLAC}
\mbox{N.B. Sinev\unskip,\iOREG}
\mbox{S.R. Smith\unskip,\iSLAC}
\mbox{M. B. Smy\unskip,\iCSU}
\mbox{J.A. Snyder\unskip,\iYALE}
\mbox{H. Staengle\unskip,\iCSU}
\mbox{A. Stahl\unskip,\iSLAC}
\mbox{P. Stamer\unskip,\iRUTG}
\mbox{R. Steiner\unskip,\iADEL}
\mbox{H. Steiner\unskip,\iLBL}
\mbox{M.G. Strauss\unskip,\iMASS}
\mbox{D. Su\unskip,\iSLAC}
\mbox{F. Suekane\unskip,\iTOHO}
\mbox{A. Sugiyama\unskip,\iNAGO}
\mbox{S. Suzuki\unskip,\iNAGO}
\mbox{M. Swartz\unskip,\iSLAC}
\mbox{A. Szumilo\unskip,\iWASH}
\mbox{T. Takahashi\unskip,\iSLAC}
\mbox{F.E. Taylor\unskip,\iMIT}
\mbox{J. Thom\unskip,\iSLAC}
\mbox{E. Torrence\unskip,\iMIT}
\mbox{N. K. Toumbas\unskip,\iSLAC}
\mbox{A.I. Trandafir\unskip,\iMASS}
\mbox{J.D. Turk\unskip,\iYALE}
\mbox{T. Usher\unskip,\iSLAC}
\mbox{C. Vannini\unskip,\iPISA}
\mbox{J. Va'vra\unskip,\iSLAC}
\mbox{E. Vella\unskip,\iSLAC}
\mbox{J.P. Venuti\unskip,\iVAND}
\mbox{R. Verdier\unskip,\iMIT}
\mbox{P.G. Verdini\unskip,\iPISA}
\mbox{S.R. Wagner\unskip,\iSLAC}
\mbox{D. L. Wagner\unskip,\iCOLO}
\mbox{A.P. Waite\unskip,\iSLAC}
\mbox{Walston, S.\unskip,\iOREG}
\mbox{J.Wang\unskip,\iSLAC}
\mbox{C. Ward\unskip,\iBRUN}
\mbox{S.J. Watts\unskip,\iBRUN}
\mbox{A.W. Weidemann\unskip,\iTENN}
\mbox{E. R. Weiss\unskip,\iWASH}
\mbox{J.S. Whitaker\unskip,\iBU}
\mbox{S.L. White\unskip,\iTENN}
\mbox{F.J. Wickens\unskip,\iRAL}
\mbox{B. Williams\unskip,\iCOLO}
\mbox{D.C. Williams\unskip,\iMIT}
\mbox{S.H. Williams\unskip,\iSLAC}
\mbox{S. Willocq\unskip,\iSLAC}
\mbox{R.J. Wilson\unskip,\iCSU}
\mbox{W.J. Wisniewski\unskip,\iSLAC}
\mbox{J. L. Wittlin\unskip,\iMASS}
\mbox{M. Woods\unskip,\iSLAC}
\mbox{G.B. Word\unskip,\iVAND}
\mbox{T.R. Wright\unskip,\iWISC}
\mbox{J. Wyss\unskip,\iPADO}
\mbox{R.K. Yamamoto\unskip,\iMIT}
\mbox{J.M. Yamartino\unskip,\iMIT}
\mbox{X. Yang\unskip,\iOREG}
\mbox{J. Yashima\unskip,\iTOHO}
\mbox{S.J. Yellin\unskip,\iUCSB}
\mbox{C.C. Young\unskip,\iSLAC}
\mbox{H. Yuta\unskip,\iAOMORI}
\mbox{G. Zapalac\unskip,\iWISC}
\mbox{R.W. Zdarko\unskip,\iSLAC}
\mbox{J. Zhou\unskip.\iOREG}

\it
  \vskip \baselineskip                   
  \baselineskip=.75\baselineskip   
\iADEL
  Adelphi University,
  South Avenue, Garden City, NY  11530 \break
\iAOMORI
  Aomori University,
  2-3-1 Kohata, Aomori City, 030 Japan \break
\iBOLO
  INFN Sezione di Bologna,
  Via Irnerio 46, I-40126 Bologna, Italy \break
\iBRUN
  Brunel University,
  Uxbridge, Middlesex, UB8 3PH United Kingdom \break
\iBU
  Boston University,
  590 Commonwealth Ave., Boston, MA  02215 \break
\iCINC
  University of Cincinnati,
  Cincinnati, OH  45221 \break
\iCOLO
  University of Colorado,
  Campus Box 390, Boulder, CO  80309 \break
\iCOLU
  Columbia University,
  Nevis Laboratories, P.O. Box 137, Irvington, NY  10533 \break
\iCSU
  Colorado State University,
  Ft. Collins, CO  80523 \break
\iFERR
  INFN Sezione di Ferrara,
  Via Paradiso 12, I-44100 Ferrara, Italy \break
\iFRAS
  Lab. Nazionali di Frascati,
  Casella Postale 13, I-00044 Frascati, Italy \break
\iILLI
  University of Illinois,
  1110 West Green St., Urbana, IL  61801 \break
\iLBL
  Lawrence Berkeley Laboratory,
  Dept. of Physics, 50B-5211 University of California, Berkeley, CA  94720 \break
\iLTU
  Louisiana Technical University,
  Ruston, LA  71272 \break
\iMASS
  University of Massachusetts,
  Amherst, MA  01003 \break
\iMISSI
  University of Mississippi,
  University, MS  38677 \break
\iMIT
  Massachusetts Institute of Technology,
  77 Massachussetts Avenue, Cambridge, MA  02139 \break
\iMOSCOW
  Moscow State University,
  Institute of Nuclear Physics, 119899 Moscow, Russia \break
\iNAGO
  Nagoya University,
  Nagoya 464, Japan \break
\iOREG
  University of Oregon,
  Department of Physics, Eugene, OR  97403 \break
\iOXF
  Oxford University,
  Oxford, OX1 3RH, United Kingdom \break
\iPADO
  Universita di Padova,
  Via F. Marzolo 8, I-35100 Padova, Italy \break
\iPERU
  Universita di Perugia, Sezione INFN,
  Via A. Pascoli, I-06100 Perugia, Italy \break
\iPISA
  INFN, Sezione di Pisa,
  Via Livornese 582/AS, Piero a Grado, I-56010 Pisa, Italy \break
\iRAL
  Rutherford Appleton Laboratory,
  Chilton, Didcot, Oxon, OX11 0QX United Kingdom \break
\iRUTG
  Rutgers University,
  Serin Physics Labs., Piscataway, NJ  08855 \break
\iSLAC
  Stanford Linear Accelerator Center,
  2575 Sand Hill Road, Menlo Park, CA  94025 \break
\iSOGA
  Sogang University,
  Ricci Hall, Seoul, Korea \break
\iSOONG
  Soongsil University,
  Seoul, Korea 156-743 \break
\iTENN
  University of Tennessee,
  401 A.H. Nielsen Physics Blg., Knoxville, TN  37996 \break
\iTOHO
  Tohoku University,
  Bubble Chamber Lab., Aramaki, Sendai 980, Japan \break
\iUCSB
  U.C. Santa Barbara,
  3019 Broida Hall, Santa Barbara,,CA ,93106 \break
\iUCSC
  U.C. Santa Cruz,
  Santa Cruz,,CA ,95064 \break
\iVAND
  Vanderbilt University,
  Stevenson Center, P.O.Box 1807, Station B, Nashville, TN  37235 \break
\iWASH
  University of Washington,
  Seattle, WA  98105 \break
\iWISC
  University of Wisconsin,
  1150 University Avenue, Madison, WI  53706 \break
\iYALE
  Yale University,
  5th Floor Gibbs Lab., P.O.Box 208121, New Haven, CT  06520  \break
\rm
%

\end{center}

\hfill
\end{document}